\documentclass[sigconf, authordraft, review=false]{acmart}

\AtBeginDocument{%
  \providecommand\BibTeX{{%
    \normalfont B\kern-0.5em{\scshape i\kern-0.25em b}\kern-0.8em\TeX}}}

\setcopyright{acmcopyright}
\copyrightyear{2018}
\acmYear{2018}
\acmDOI{XXXXXXX.XXXXXXX}

\acmConference[Conference acronym 'XX]{Make sure to enter the correct
  conference title from your rights confirmation email}{June 03--05,
  2018}{Woodstock, NY}
\acmPrice{15.00}
\acmISBN{978-1-4503-XXXX-X/18/06}

\usepackage[nolist,nohyperlinks, printonlyused, withpage]{acronym}
\usepackage{csquotes}
\usepackage{siunitx} 
\usepackage{tabularx}
\usepackage[inline]{enumitem} 

\usepackage{subcaption}
\usepackage{natbib}
\usepackage{soul}
\usepackage{fontenc}
\usepackage[resetlabels]{multibib}
\newcites{game}{Ludography}
\newcommand{\citegameprefix}{G}
\usepackage{xparse}
\let\origcitegame\citegame
\RenewDocumentCommand{\citegame}{O{} O{} m}{%
  \renewcommand{\citenumfont}[1]{\citegameprefix##1}%
  \origcitegame[#1][#2]{#3}%
  \renewcommand{\citenumfont}[1]{##1}%
}
\newcommand{\citeg}[1]{\citegame{#1}} 

\SetBlockEnvironment{italicquote}

\DeclareQuoteStyle{italic}[\itshape][\itshape]
{\textquotedblleft}{\textquotedblright}
{\textquoteleft}{\textquoteright}
\setquotestyle{italic}

\newcommand{\tdf}[2]{$t_{#1}{=}#2$}

\newcommand{\psig}[2][=]{$p{#1}#2$}
\newcommand{\n}[1]{$n{=}#1$}
\newcommand{\cohensD}[1]{Cohen's $d{=}#1$}
\newcommand{\MeanSd}[2]{$M{=}#1$, $SD{=}#2$}

\begin{document}



\title[Disengagement from Games]{Disengagement From Games: Characterizing the Experience and Process of Exiting Play Sessions}

\author{Dmitry Alexandrovsky}
\authornote{Both authors contributed equally to this research.}
\email{dmitry.alexandrovsky@kit.edu}

\author{Kathrin Gerling}
\authornotemark[1]
\email{kathrin.gerling@kit.edu}
\affiliation{%
  \institution{Karlsruhe Institute of Technology}
  \city{Karlsruhe}  
  \country{Germany}  
}

\author{Merlin Steven Opp}
\affiliation{%
  \institution{Karlsruhe Institute of Technology}
  \city{Karlsruhe}  
  \country{Germany}
}

\author{Christopher Benjamin Hahn}
\affiliation{%
  \institution{Karlsruhe Institute of Technology}
  \city{Karlsruhe}  
  \country{Germany}
}

\author{Max V. Birk}
\affiliation{%
  \institution{Eindhoven University of Technology}  
  \city{Eindhoven}  
  \country{Netherlands}
}

\author{Meshaiel Alsheail}
\affiliation{%
  \institution{Karlsruhe Institute of Technology}
  \city{Karlsruhe}  
  \country{Germany}
}

\renewcommand{\shortauthors}{Alexandrovsky and Gerling, et al.}

\begin{acronym}
\acro{ARCS}{attention, relevance, confidence, satisfaction}
\acro{CEGE}{Core Elements of game experience}
\acro{DDA}{Dynamic Difficulty Adjustment}
\acro{FUGA}{Fun of gaming}
\acro{FPS}{First-person shooter}
\acro{GUR}{Games User Research}
\acro{HCI}{Human-Computer Interaction}
\acro{KIT}{Karlsruhe Institute of Technology}
\acro{MMO}{Massively Multiplayer Online}
\acro{MMORPG}{Massively Multiplayer Online Role-playing Game}
\acro{PC}{Personal Computer}
\acro{PE}{Player Engagement}
\acro{PX}{Player Experience}
\acro{PXI}{Player Experience Inventory}
\acro{RPG}{Role-playing Game}
\acro{SDT}{Self-Determination Theory}
\acro{TAM}{Technology Acceptance Model}
\acro{UE}{User Engagement}
\acro{UX}{User Experience}
\acro{VR}{Virtual Reality}
\end{acronym}

\begin{abstract}
The games research community has developed substantial knowledge on designing engaging experiences that draw players in. Surprisingly, less is known about player \textit{dis}engagement, with existing work predominantly addressing disengagement from the perspective of problematic play, and research exploring player disengagement from a constructive designer perspective is lacking. In this paper, we address this gap and argue that disengagement from games should be constructively designed, allowing players to exit play sessions in a self-determined way. Following a two-phase research approach that combines an interview study (\n{16}) with a follow-up online survey (\n{111}), we systematically analyze player perspectives on exiting play sessions. Our work expands the existing notion of disengagement through a characterization of exit experiences, a lens on disengagement as a process, and points for reflection for the design of games that seek to address player disengagement in a constructive way. 

\end{abstract}

\ccsdesc[500]{Human-centered computing~HCI theory, concepts and models}
\ccsdesc[500]{Human-centered computing~Empirical studies in HCI}

\keywords{Disengagement, Exiting Games, Player Experience}

\received{20 February 2007}
\received[revised]{12 March 2009}
\received[accepted]{5 June 2009}

\maketitle

\section{Introduction}
Game design aspires to engage players, employing design strategies that draw players in and keep them within the game. Previous work within the \ac{HCI} games research community has, for example, examined the relevance of player onboarding for continued player engagement~\cite{cheung2014}, how player interaction patterns such as snacking can be leveraged to combat player attrition~\cite{alexandrovsky2019}, or how we can measure constructs relevant in the context of engagement~\cite{abeele2020, jennett2008, johnson2018}. Overall, there is a strong emphasis on building and retaining engagement, and our community has built a comprehensive body of knowledge thereof. In contrast, a more recent body of \ac{HCI} games research has begun to explore player experiences beyond this mainstream, e.g., \citet{tyack2021} report that many players do not seek highly intensive experiences but rather enjoy playing with low tension for relaxing. However, less is known about the experience of \textit{dis}engagement or the process and experience of reducing involvement and exiting a gaming session. Here, existing work on casual games suggests that the final moments of play are highly relevant for the overall player experience~\cite{gutwin2016}, and another body of literature addresses disengagement through the lens of problematic play (e.g., obsessive engagement~\cite{przybylski2009a} and behavioral game design that is coercive and promotes ongoing engagement ~\cite{zagal2013}). 

However, there is a gap in scholarship that explores player disengagement from a qualitative perspective: Little is known about how players experience the end of a play session, and it is not clear what characterizes player disengagement, which we understand as the process of ending a play session with the intention of resuming play at a later point in time. This indicates a need to understand how we account for disengagement in game design, treating it as a regular part of play that has the potential to contribute to a positive player experience. Thus, our work seeks to address this gap through the following two main research questions (RQs):

\begin{enumerate}
    \item[\textbf{RQ1:}] How do players experience disengagement from a play session, and what is considered a \textit{positive} exit experience? 
    \item[\textbf{RQ2:}] How do players structure the process of disengaging from a play session, and which characteristics of games help or hinder this process?
\end{enumerate}

In our work, we approach player disengagement on the basis of ~\citeauthor{obrien2008}'s model of \acf{UE}, which conceptualizes engagement as a cyclical process (the Engagement Cycle) in four phases with varying intensities of experience~\cite{obrien2008} (see Section~\ref{subsec:engagement_cycle}). Building upon the Engagement Cycle, we establish an exploratory two-phase research process that first provides in-depth insights into individual exit experiences through exploratory interviews and then examines disengagement across a broader group of players in the context of an online survey that builds upon main interview findings. This methodological approach allows us to generate insights within an under-explored area of research~\cite{creswell2017} while producing rich data conducive to our exploratory lens~\cite{braun2021} and to characterize users' attitudes, intents, and experiences when interacting with technology ~\cite{lazar2017, muller2014}. 

Results of semi-structured interviews with $16$ participants show that participants perceive exit experiences as positive when they left the game with a sense of satisfaction and closure, and that they find it easiest to do so in games that allow them to easily understand the structure of play, and that are transparent in terms of exit options. At the same time, participants were acutely aware of instances in which games hampered their ability to disengage with agency, which is something to be taken into account by game design. Overall, we also observed that participants were strategic in how they approached disengagement, i.e., many of them set goals for the duration of and achievements within play, and leveraged their knowledge of the structure of games in the process. 

Survey responses from $111$ participants support the main interview findings, showing that successful disengagement is associated with closure, and carried out with agency. Additionally, findings highlight the complexity of disengagement experiences, suggesting that disengagement can be emotionally challenging for players while still being perceived as positive and that disengagement experiences do not only fall into either positive or negative categories but can also be neutral, expanding upon existing work on user engagement~\cite{obrien2013,obrien2016,obrien2022}.
Additionally, survey results support the notion that players actively consider and plan for disengagement, but add more nuance to the context in which this takes place: Our findings suggest that disengagement is impacted by external factors, for example, players feeling tired or needing to engage with other tasks, or needing to attend to their social life.

Our work makes the following main contributions: 
\begin{enumerate*}[label=(\arabic*)]
    \item We provide the first exploration of players' experiences of disengagement as a regular part of engaging with games, and we characterize positive, neutral, and negative disengagement experiences.
    \item We provide points for reflection for researchers and designers wishing to create positive exit experiences, and we highlight how individual approaches to the process of disengagement and player context need to be taken into account to facilitate self-determined disengagement.
    \item We discuss the need to engage players while also allowing for disengagement, and we highlight future opportunities for our research community in the context of player disengagement.
\end{enumerate*}

\section{Related Work}
In this section, we draw from literature in \ac{HCI} and games research to discuss how disengagement in games is currently conceptualized. We also introduce the Engagement Cycle as the theoretical foundation for our work.

\subsection{Engagement and Disengagement in Games}
\label{subsec:disengagement_in_games}
Games are interactive systems inherently designed to be enjoyable~\cite{deci2000, gee2007, rigby2011, vorderer2004, yee2006}.
Hence, the primary goal of game design focuses on providing players with more enjoyable experiences, trying to capture them in the game for as long as possible~\cite{boyle2012, deterding2023, hamari2015, rapp2022, soderman2021}. Over the last decade,~\ac{GUR} has formalized game design and developed a substantial understanding of components in games that provide enjoyable experiences in games (\ac{PX})~\cite{fullerton2014, nacke2017, salen2004, schell2020}, which is defined as \textquote{the qualities of the player-game interaction and is typically investigated during and after the interaction with games}{~\cite{wiemeyer2016}}. 
\ac{PX} derives from \ac{UX} and takes motivation~\cite{birk2016, koivisto2019}, emotion~\cite{bateman2006,bouvier2014, hallifax2019}, and personality~\cite{nacke2014, teng2008,zammitto2010} into account. Harnessing player enjoyment of games and the desire to play, game design has been widely studied and applied to motivate people to engage in activities they otherwise find difficult to do, such as exercising~\cite{kappen2019}, learning~\cite{gee2007, prensky2003}, or adhering to a therapy program~\cite{johnson2016,neupane2021}. In fact, one of the most cited pieces of work within the wider games research community extensively discusses the motivational pull of games on the basis of Self-Determination Theory \cite{ryan2006}, suggesting that games engage players because they have the potential to fulfill the basic human needs of autonomy, competence, and relatedness. 

\subsubsection{Disengagement as the Absence of Engagement}
\label{subsubsec:disengagement_absence}
Consequently, games that are boring or yield negative emotions, such as anger or frustration, are often understood to be poorly designed~\cite{crawford1984, koster2013}. This also manifests in the focus of game research on the design of game elements that facilitate an engaging experience. For example, ~\ac{DDA}~\cite{hunicke2005} is a widely applied design paradigm to enhance perceived competence and avoid boredom as well as frustration while maximizing the Flow experience~\cite{chanel2008, bowman2021}. The underlying assumption is that balancing the game challenges with the players' skills facilitates the experience of Flow, providing enjoyment and reducing drop-out~\cite{mekler2014}. Similarly, today's games often implement visual embellishments -- juiciness~\cite{schell2020} -- promoting the overall aesthetic appeal and pleasurable experiences~\cite{hicks2019,bateman2010, juul2010}.
In this sense, design for engagement is frequently operationalized as adherence~\cite{lumsden2016, kelders2012, weber2011}, i.e., how often and how many users/players visit the respective platform~\cite{fields2011, gjoka2008, mendelson2018}. Accordingly, research to date tends to conceptualize disengagement or non-adherence as the absence of engagement. However, how much players adhere to the game is often disentangled from whether they enjoy it~\cite{nacke2011} or are motivated to play~\cite{mekler2013, mekler2017, jolley2006}. Such designs often draw players into the \textit{vortex}~\cite{moran2018} -- a design pattern where users are pulled towards unplanned interactions and lose the sense of control -- that can foster an obsessive passion~\cite{johnson2021} for interaction with (playful) media, with negative consequences for the individuals' well-being~\cite{luxford2022}. 
This aligns with a wider challenge to address unethical design practices~\cite{mildner2021} that are dis-empowering players~\cite{vornhagen2023}, lever out one's self-regulation mechanisms~\cite{alter2018, moran2018, howe2017}. Such \textit{deceptive} design patterns~\cite{zagal2013} do not let users focus on concurrent stimuli or engage with other activities meaningfully~\cite{atkinson2006, fogg2002}. 
This negative impact of behavioral design is increasingly recognized in the games research community (e.g.,~\citet{birk2023, zagal2013}).

Overall, this suggests that there is an opportunity for games research to contribute to the design of games that facilitate meaningful engagement, but also adopt a broader lens on disengagement. Here, it is helpful to understand how \ac{HCI} games research has explicitly addressed player disengagement to date.


\subsubsection{Disengagement as Unique Part of the Player Experience}
\label{subsubsec:disengagement_unique}
Games research has previously addressed player disengagement from different perspectives. Adopting a designerly approach,~\citet{bjork2004} describe design patterns for closure to structure the progression of games, implicitly touching upon disengagement, but not providing an empirical exploration thereof. Likewise, general design recommendations of narrative structures and level design argue for structures that facilitate reduced engagement through the provision of tension directly followed by phases of release so that players can reflect on the intense experience and recover~\cite{kotler2014}. For instance, players enjoy games as a means to recover from daily activities such as work~\cite{collins2014}, and often players engage with video games \textit{ordinarily} to relax, not seeking exceptional experiences~\cite{tyack2021}. 
However, while these pieces of work do consider reduced or low-key engagement, they do not expand to disengagement. One example of work within the \ac{HCI} games research community that does address disengagement constructively is provided by~\citet{gutwin2016}, who manipulated the degree of challenge in a sequence play rounds and confirmed the Peak-End rule (i.e., that peaks and the end of an experience substantially determine the overall evaluation thereof) for casual games. They show that players recall sequences with positive peaks and endings as more enjoyable and interesting and are more willing to re-engage over sequences with neutral and negative peak ends. However, the study does not provide insights into the structure and the process of disengagement from the perspective of the player, and restricts itself to performance and challenge the main determinant of \ac{PX}. 
Finally, adjacent work by~\citet{knibbe2018} explores the moment of disengaging from \ac{VR}, seeking to understand how people experience the transition from the virtual world back into reality. While this work strongly focuses on the element of sensory immersion in \ac{VR} and therefore has limited implications for disengagement from games more generally, it does highlight the importance of understanding the exit experience when designing for engaging interaction.

\subsubsection{Existing Tools and Strategies to Facilitate Disengagement}
\label{subsubsec:disengagement_tools}
Beyond work that addresses player experience, there have been some efforts to address player disengagement through the lens of problematic engagement~\cite{ferguson2007}, where games are canonized as inherently addictive~\cite{bean2017, elsayed2021, kuss2013} and violent~\cite{coyne2018, griffiths1999}.
Hence, research and industry have explored strategies to reduce interaction time by introducing tools and strategies to facilitate disengagement, most commonly through the introduction of time restrictions, either at an individual or societal level.
Such tools involve timers~\cite{hiniker2016a}, trackers of usage~\cite{kim2016a, fioca2007}, automated nudges to disengage~\cite{okeke2018}, promoting self-regulation through social support and goal-setting~\cite{ko2015}, or block users entirely from using the device or specific apps~\cite{lee2014, jasper2015}. Specifically for children, various tools seek to manage screen time and other issues by allowing parents to set time limits~\cite{bieke2016}, e.g., Net Nanny~\cite{ross2021}, CYBERsitter~\cite{milburn1998}, or child-friendly filters on Netflix and Apple's ScreenTime\footnote{\url{https://support.apple.com/en-ca/HT208982}} which also provide detailed statistics about the usage of individual apps. Likewise, there are less authoritative approaches to the management of playing time. For example, console platforms implement functionality that allows players to check time spent on a current session or let players configure individual reminders to take breaks (e.g., see~\cite{xboxsupport}).
Similarly, casual and free-to-play games exploit time restriction (e.g., as a resource) as part of their monetization model~\cite{fields2011,xiao2021,xiao2021a}.
\citet{alexandrovsky2019, alexandrovsky2021} describe these game elements \textit{waiting} and \textit{blocking} mechanics, albeit not with the intention of facilitating player-led disengagement. However, time-based approaches to disengagement do not account for player experience throughout the process (e.g., the benefit of competence~\cite{gutwin2016, ryan2006}), and are a missed opportunity for game design to support disengagement.

Overall, this suggests that there is a gap in \ac{HCI} games research that approaches disengagement from a broad and constructive perspective, viewing it as a natural part of play rather than problematizing it. In our work, we seek to address this issue through an initial exploration of the ways players structure and experience disengagement from the games they regularly play.

\subsection{The Engagement Cycle}
\label{subsec:engagement_cycle}
Expanding on the notion of \ac{UX} and \ac{PX}, 
\citet{obrien2008} propose the concept of \ac{UE}, which they define as~\textquote{a quality of UX that is characterized by the depth of the actor’s investment in the interaction; this investment may be defined temporally, emotionally, and/or cognitively}{~\cite{obrien2016}.} On this basis, they contribute a model of engagement that conceptualizes interaction over time in the \textit{Engagement Cycle} consisting of four stages:
\begin{enumerate*}[label=(\Roman*)]
    \item the \textit{point of engagement} is the first contact with the interactive system,
    \item the \textit{period of sustained engagement} is the actual time span users interacting with the system,
    \item \textit{disengagement} describes the termination point of an engaging period (e.g., end of a session), and
    \item \textit{re-engagement} is referred to when users return to the interactive system at their own volition
\end{enumerate*}.

In the context of our research, such a cyclical model is helpful in understanding player experience across different stages of play. This view on interaction with media acknowledges that it is a process associated with different experiences and explicitly includes disengagement.
Reflecting on harmonious and obsessive engagement, the model of \ac{UE} describes both positive and negative forms of \textit{en}gagement on the two dimensions of \textit{value} and users' \textit{agency}~\cite{obrien2022}. These two dimensions  of value and agency correspondingly align with how we approach player experience, i.e., the understanding that games should provide enriching experiences to players and that player autonomy or agency is a central element of positive~\ac{PX}~\cite{deterding2016, ryan2006}.
However, \citet{obrien2022} are less specific when it comes to the experience of disengagement, where the engagement cycle depicts disengagement as a natural part of long-term interaction in the users' daily life, where engagement is not continuous but instead fluctuates. 
With this, positive engagement and disengagement are associated with the users' achievement of their goals and their agency in engaging. Therefore, to support positive engagement and to account for individual user needs, \citeauthor{obrien2022} suggest natural breakpoints in the interaction design \textquote{such as recommending additional resources, stopping at the end of a module, or letting users choose the length of an interaction}{~\cite{obrien2022}}.
The cursory treatment of disengagement is also a weakness of competing models such as the OA3 framework~\citet{schoenau-fog2011}, which describes \ac{PE} as a broader construct that conceives play as a process focused on \textit{sustained engagement} or ~\textquote{what makes people want to continue playing}~{\cite{schoenau-fog2011}}. 
With respect to disengagement, the authors specifically investigate triggers of engagement and disengagement but consider disengagement to be negative. In this context, disengagement is conceptualized as lacking engagement, such as a mismatching challenge, players getting stuck, bored, or not being satisfied with in-game content. 


In the context of our work, we draw upon existing considerations regarding (dis)engagement to shape our qualitative exploration of how players experience the end of a play session. Specifically, we build upon the cyclical nature of the Engagement Cycle and set out to explore positive and negative disengagement experiences along with the process of disengagement, which is in line with~\citet{obrien2022}'s binary conceptualization of \ac{UE}. This approach allows us to address disengagement from the position that neither assumes it to be inherently negative nor positive.

\section{Phase 1: Interview Study}
\label{sec:interviews}
In the first phase of our work, we employed semi-structured interviews (\n{16}) to explore in depth how players experience disengagement from games. In particular, we focused on experiences that players had previously made and the strategies they employed when exiting a play session.

\subsection{Method}
Participants were invited to take part in semi-structured interviews that examined how they experienced disengagement from games. We opted for a qualitative research approach as this is well-suited to capture lived experience~\cite{braun2013}. 
For this part of our research, we recruited participants who considered gaming a regular and valued form of leisure but did not have a history of problematic gaming. We also attempted to recruit a participant sample with diversity regarding gender and age. We did not apply any further inclusion or exclusion criteria in the recruitment process. 
The interview questions were designed against the backdrop of the Engagement Cycle~\cite{obrien2016, obrien2022}, and in a fashion that neither assumed exclusively positive or negative disengagement experiences. First, we asked participants to reflect upon exit experiences in relationship with games they frequently play, e.g., \textit{"What was your experience like when you left the game?"}, seeking to explore the different experiences people had when leaving play. Then, we asked participants to describe their behavior when exiting a play session, e.g., \textit{"How and when do you decide to exit a play session?", and "What strategies do you practically employ when leaving a play session?"}. We also explored whether disengagement had a relationship with re-engagement (cf. Section~\ref{subsec:engagement_cycle}). Then, we examined whether participants found particular characteristics of games helpful in this process, e.g., \textit{"Can you recall any characteristics of games or game elements that helped you end a play session?", "How did these elements impact the experience that you had with the game?"}. Finally, we asked players about their own perspectives on what could facilitate a positive exit from play, and we collected demographic information, including gaming habits as well as gaming preferences. For the full interview guide, please refer to the supplementary materials.

\subsection{Procedure}
We recruited participants via social media channels (Twitter, Facebook, LinkedIn, and Mastodon) and word of mouth between May 2023 and June 2023. 
Depending on the participants' preference, interviews either took place in person at a preferred venue or online. 
On average, interviews lasted about $30$ minutes. 
At the beginning of each interview, participants were informed about the research, given space to ask questions about the procedure, and were asked to provide informed consent. 
Afterward, we gathered demographic information and then followed up with their experiences when exiting games. At the end of each session, participants were given the opportunity to ask additional questions. Interviews were conducted in English and German and transcribed immediately after they were held. The research protocol was approved by the <removed for review> ethics board.

\subsection{Participants}
Sixteen participants (nine men, six women, one non-binary person; average age $27.5$ years, range $20{-}47$) took part in the interviews. 
All but one participant resided in Western Europe, living in a range of arrangements, e.g., with roommates, a partner, or family. The average playing time per week ranged from two to $45$ hours, with an average of $13.06$ hours, suggesting that participants regularly and extensively engaged with games. 
The average number of days played per week ranged from one to seven, with an average of $4.56$. The average playing time per session was $2.39$ hours and ranged from $1.5$ to $3$ hours. 
The majority of participants reported playing on \ac{PC} and console platforms (e.g., Nintendo Switch, Sony PlayStation, Microsoft Xbox), and a smaller number of participants regularly played \ac{VR} games. 
Likewise, some participants reported playing mobile games. 
In terms of favorite and regularly played games, there was a breadth of interests in our participant sample, ranging from \acp{RPG} (e.g., the Diablo series~\citeg{blizzardentertainment2023}) to \ac{FPS} games (e.g., the Call of Duty series~\citeg{infinityward2022}) and sports titles (e.g., FIFA~\citeg{eavancouver2022}) to cozy games (e.g., Stardew Valley~\citeg{concernedape2016}) and a range of indie games. 
With respect to their relationship with games, all participants reported playing games as a hobby, with a small number of participants also engaging with games in the context of their jobs. 
Within our sample, no participant reported significant difficulty when disengaging from play; when difficulties disengaging were reported, these were typically related to one specific game that players perceived as highly engaging. However, beyond feeling tired the next day, participants reported no negative implications on their lives.

\subsection{Data Analysis}
Data were analyzed using an inductive approach, applying Reflexive Thematic Analysis~\cite{braun2012}, which is an analysis method that is well-suited when examining participants' experiences, thoughts, and behaviors in the context of qualitative research~\cite{kiger2020}. 
Following the process laid out by the authors, a member of the research team first familiarized themselves with the data through reading and re-reading the transcripts and then engaged in an initial coding process and the shaping of an initial set of themes. 
The process of crafting themes was supported by other members of the research team, who engaged in discussion throughout and contributed to a joint review of the themes.

\subsection{Positionality}
Given that we engaged in an interpretative, analytical approach, we make the positionality of our research team explicit. 
Our team comprises various academic backgrounds, ranging from computer science to psychology; one member of the research team previously worked in the games industry. 
Overall, there were a range of perspectives on and experiences with disengagement from games: All members of the research team played games at some point in their lives; one member of the research team is a parent who now also experiences disengagement from play through the experience of their child. Some members of the team have individual lived experiences of finding it difficult to disengage, while others do not share this experience; one member of the research team no longer plays games because disengagement remains an issue. Generally, the team views gaming as a positive activity and none that needs to be artificially regulated. 
However, there also is agreement that not all strategies employed in game design are fair for players (e.g., those that exploit human psychology), and that player self-determination should be facilitated through game design that prioritizes supporting players in their own decisions of when to exit play.

\subsection{Results}
In this section, we present the three main themes that were crafted during the analysis,
\begin{enumerate*}[label=(\arabic*)]
    \item[\nameref{subsubsec:interview_theme_1}:] Positive disengagement experiences are associated with satisfaction and closure, while negative disengagement experiences come with feelings of rage and regret,
    \item[\nameref{subsubsec:interview_theme_2}:] disengagement is a complex process that spans players, games, and their environment, and
    \item[\nameref{subsubsec:interview_theme_3}:] players' agency is a central element of positive disengagement
\end{enumerate*}.

\subsubsection[Theme 1]{Theme 1: Positive Disengagement Experiences Come With Satisfaction and Closure, Negative Ones With Rage and Regret} \label{subsubsec:interview_theme_1}
Players had a range of experiences when exiting games, with the majority of positive experiences centering around these that left players with a sense of satisfaction and took place at a point where players perceived that they had reached a point of closure within the game. For example, one participant noted that \textquote[P09]{\textelp{} in Celeste, when you wrap up a level or a part of a level, then you feel good, and that can last a while, and that's a good time to stop \textins{playing}}{,} and another one noted that \textquote[P11]{\textins{after in-game success} it is like, I'll wrap up for the day, I've had enough, and that is very satisfying}{.} 
In this context, participant feedback revealed that the achievement of closure is often experienced as cyclical, can involve multiple strands of action, and is characterized by players achieving their goal in the context of the game, e.g., finishing a section of the game or accomplishing a certain task.


With respect to what contributes to a \textit{satisfying} experience, participants offered a range of perspectives on perceived benefits of play, which were often associated with progress leading to a specific achievement. For example, one participant noted that \textquote[P11]{sometimes you already know, okay, I'll need this many points, and then you work towards that goal. \textelp{} And then I carry out that one activity, and then I end the game}{.} Likewise, many participants reported overcoming significant in-game challenges (e.g., beating a final boss or winning a match in multiplayer settings) as a source of satisfaction; interestingly, we did not observe instances where achievement was explicitly associated with other kinds of benefits. However, we also note that players aligned their choice of games with a wider range of anticipated benefits, with one participant noting that \textquote[P15]{\textins{i}f I only have one-hour play session available in my night or two-hour play session I’m going to choose to bond with \textelp{} my partner and play Overwatch with him rather than playing \textelp{} Plague Tale by myself}, suggesting that social experiences may also provide satisfaction. 

Likewise, in experiences where progress and achievement fell short of what was expected, disengagement was associated with disappointment. In extreme cases, we do note a number of distinctly negative accounts of exiting play that fall into this category. 
Most prominently, repeatedly performing poorly in multiplayer matches was reported as a reason for unpleasant disengagement associated with anger, e.g., with one participant pointing out that \textquote[P14]{I stopped playing, out of frustration because I couldn't think clearly anymore... because I was angry.}. Along the same lines, participants commented that they had previously quit play sessions early and in frustration as a result of buggy games, e.g., \textquote[P01]{\textelp{} The game crashed multiple times. And then I simply quit playing because I was really frustrated that I had to keep playing the same missions over and over again}{.} Likewise, there were participants who quit games because they were bored, i.e., because the games did not provide the experience they expected. In some of these situations, participants expressed regret over time that they, in the end, did not consider well spent, or feelings of guilt when playing longer than planned, e.g., \textquote[P05]{Yeah, I do feel a little guilty \textins{when disengaging later than planned}, because, yeah}{.}



\subsubsection[Theme 2]{Theme 2: Disengagement is a Complex Process Involving Players, Games, and the Environment}
\label{subsubsec:interview_theme_2}
This theme highlights player approaches to structuring disengagement from games. Our results highlight that disengagement is a complex process that cannot be reduced to one moment in time at which a player decides that it is time to stop playing. Instead, it is a multi-step process involving several phases and taking into account the state of the player, the game, and other environmental factors. In terms of phases, many participants reported pre-play disengagement plans, i.e., choosing a game that fits the amount of time available, and making decisions about the duration of play before engaging with the game, for example, \textquote[P09]{If I do have experience with the game, then I use it, to say, well, I'll manage three runs and then I will stop}{.} Here, we observed that time constraints often went along with in-game goals, e.g., planning on finishing a level or mission within a game. This implies that players enter the play session with a predefined moment in time - or goal - at which they wish to disengage. 

However, participant feedback also revealed that this goal can fluctuate in accordance with the state of the game, suggesting an in-game adjustment of disengagement decision. 
For example, one participant highlighted that they would consider exiting \textquote[P10]{in a quiet moment of play, when things aren't so exciting, then I would quit}; \textquote[P12]{not in a moment where I still need to remember multiple things \textins{related to the game}}, suggesting an adjustment to the flow of gameplay and preference for in-game situations with low cognitive load. 
Likewise, participants also recognized that disengagement was sometimes more difficult when fully engaged in the game, e.g., \textquote[P03]{\textelp{} once you are in a topic, then it is undesirable to stop, once you're in the flow}, suggesting that pre-play plans can be hard to execute during play. This was closely related to a strong desire to exit at a point of closure, which positively impacted the ability to execute a disengagement decision, e.g., \textquote[P03]{\textelp{} the end of the level or a task is a good point to stop; otherwise it'll be an endless loop, I mean, when would you quit?}. Here, some participants also reported a feeling of cognitive exhaustion as a contributor to disengagement, e.g., \textquote[P02]{At some point, I'm simply exhausted. And then I don't feel like playing anymore}, or -- in the case of \ac{VR} gaming -- physical exhaustion: \textquote[P03]{\textins{When playing Beat Saber} I often start to feel hot, I sweat, and then I stop at some point}. Players also reported the impact of external factors on their decision to end play. For example, participants pointed out that other responsibilities or goals that were not game-related were a reason to decide to end play sessions in situ, e.g., chores, studies, partners and friends coming home, or the end of a commute, e.g., \textquote[P16]{So when my train is at the stop, I need to get off. Then, you know, obviously, I need to stop playing because otherwise I miss my stop and I don't get home}{.}

Finally, there were some examples where players connected disengagement with considerations for re-engagement, thereby facilitating the process of ending play. For example, one participant highlighted that they were preparing for re-engagement: \textquote[P12]{I tend to keep playing until I am at a point where, yes, okay, that is fitting, and it won't be too awkward to continue from there the next time that I play}{.} 
Likewise, while the large majority of participants fell in line with the processes described above, we also observed instances in our data where the disengagement process was heavily impacted by the emotional state of the player. Instead of employing \textit{orderly} disengagement processes, some participants reported situations in which they \textit{ragequit}, i.e., situations in which they abruptly left the game in response to in-game situations that they experienced as frustrating (see Theme 1). 

\subsubsection[Theme 3]{Theme 3: Player Agency is a Central Element of Positive Disengagement}
\label{subsubsec:interview_theme_3}
The final theme is concerned with player agency, i.e., their ability to \textquote{act as they wanted}{~\cite{bopp2016},} and the role that it plays in the disengagement process. This is an aspect that was repeatedly alluded to by players, highlighting the relevance of player-led strategies to disengage from games and spanning both the experience and the process of disengagement. Here, two relevant sub-themes are the facilitating role of mental models of games when exiting play and the need for transparent communication of current and future game states to support disengagement with a view to re-engagement.

Overall, we found that players value when games enable them to make their own decisions about when to disengage, leading to a positive perception of the play session in retrospect. For example, participants extensively discussed the benefits of being able to save their progress at all points of play and highlighted the importance of games offering explicit options to leave the game (as opposed to simply switching off the console, which is popular on some platforms). In contrast, players were aware that in situations where their agency was reduced, disengagement became more difficult. For example, one participant pointed out that~\textquote[P11]{\textins{in Overwatch 2} there’s not really an exit button, and when it’s very hidden, so when you’re in the main menu. I think you either have to press Escape, or in the bottom right corner, there’s kind of an option to exit the game, but it’s not that easy to exit the game; you get sucked into it very easily for one more round}{.} Here, many players reported a negative player experience, e.g., \textquote[P03]{I felt pressured by the game \textins{not offering save points}}{.} However, difficulty with disengagement was not only associated with practical game features but could also be a result of engaging storylines that maintained player curiosity, e.g., \textquote[P05]{\textins{disengagement was difficult} because I wanted to continue with my character, to unlock new skills, and to see how they work}{.}

From the perspective of games that support player agency at the point of disengagement, participants referred to concrete game features and elements that supported their agency. For example, features such as autosaving and games that allowed players to save at any point in play were appreciated by the vast majority of participants as a means of giving security that no progress would be lost regardless of the point of disengagement while not interfering with gameplay and allowing them to \textquote[P02]{stay in the flow of the game}. Likewise, participants noted that the narrative structure of games could be a support for disengagement, pointing out that \textquote[P10]{\textelp{} where you have encapsulated threads played by characters, when you have quests, it is easier to find a point where you stop, rather than in games where threads that remain open}{.} Aligning with the desire for agency, participants pushed back against approaches that would restrict playing time, e.g., \textquote[P01]{I wouldn't appreciate it. Simply because if I have made a plan, I am going to play today, and the game tells me after an hour, thank you for playing, now you're done, then I think I would probably choose a different game next time, without that feature}{.} Interestingly, some participants also reported using external tools to support disengagement, e.g., \textquote[P02]{If I have an hour to play throughout the day, then I set a timer \textelp{}. When the timer goes off, then it gives me something to do \textins{outside of the game}, and that allows me to switch off \textins{the game}}{.}

With respect to the sub-theme of the facilitating role of mental models of games in the disengagement process, our results show that players who succeeded in creating mental models of games they play were in a position to leverage their knowledge about the structure of the game to make informed decisions about disengagement. For example, one participant pointed out that \textquote[P13]{\textins{when I the know match length}, then I say, okay, I can only play two more games, and then I'm done}{.} Along the same lines, another participant highlighted that \textquote[P07]{\textelp{} with Stardew Valley, where you have to go to sleep in order to save, I think about whether I should start a new day, because that also takes, I don’t know how long the days are, but at least 20 minutes to play the whole day}{.} This shows how a mental model of the game's structure can serve both as a tool to plan for disengagement (see Theme 1) and a prompt to reflect upon the need to close a play session. This is closely aligned with the final sub-theme, the need for transparent communication of current and future game state, which relates to establishing mental models of games and supporting agency by allowing players to make informed exit decisions. For example, many participants highlighted that insecurity about what progress would be saved at which point in the game was a cause of frustration, in some cases leading to permanent disengagement from the game, highlighting the relevance of transparent communication throughout the final stages of play.

\section{Phase 2: Online Survey}
The first phase of our research (see Section \ref{sec:interviews}) identified three initial themes associated with the following core findings: 
\begin{enumerate*}
    \item[\nameref{subsubsec:interview_theme_1}:] Positive disengagement experiences are associated with satisfaction and closure, while negative disengagement experiences come with feelings of rage and regret,
    \item[\nameref{subsubsec:interview_theme_2}:] disengagement is a complex process that spans players, games, and their environment, and
    \item[\nameref{subsubsec:interview_theme_3}:] players' agency is a central element of positive disengagement
\end{enumerate*}.
To further expand on these results, we conducted an online survey with a strong qualitative component where we asked a wider audience to describe their experiences when exiting play sessions. In particular, our online survey aimed to discuss the main themes that were crafted in the first phase of our work with a broader group of players and to provide a more nuanced perspective on the themes. We first invited participants to rate key findings from the first phase using Likert scales to stimulate reflection, and then asked them to elaborate on their responses qualitatively~\cite{walsh2013}. This is a common research approach in \ac{HCI}, where surveys are frequently leveraged to expand upon initial findings and guide further research direction~\cite{dell2016, muller2014}. Generally, surveys with qualitative components offer the opportunity to produce rich data and invite broad participation~\cite{braun2021}, supporting our exploratory research approach.

\subsection{Method}
The survey consists of three parts: First, respondents were asked to provide demographic information about themselves and their relationship with games. In the second section, we explore respondents' positive and negative experiences when they exited a play session in an open-ended way through open-ended questions, and we asked the respondents to express whether they felt a sense of \textit{achievement}, \textit{closure}, or \textit{satisfaction} on 7-point Likert scales ranging from Strongly disagree ($1$) to Strongly agree ($7$). In the third and last major section, in line with the themes derived from the interviews, respondents were asked to rate general statements about how the \textit{process of disengagement} takes place, the \textit{respondents' experience} when exiting a play session, and which \textit{game elements affect disengagement} on a total of nine 7-point Likert scales ranging from Strongly disagree ($1$) to Strongly agree ($7$). Sample items include \textit{A positive exit experience is one where I feel a sense of satisfaction}, and \textit{When I exit a play session, it is important to me to be in control of the process}. The intention for inclusion of these items was to offer respondents an entry point into the topic and to stimulate reflection, which is in line with best practices in survey application in \ac{HCI}~\cite{walsh2013} and beyond~\cite{lutz2013}. Following each item, respondents were invited to leave open-ended comments to further explain their responses. Finally, respondents could address any issues they felt were not already covered by the survey. In total, the survey was designed to be completed within $15{-}20$~\si{\minute}. The complete questionnaire is included in the supplementary materials.

We estimated the required sample size for our survey following general guidelines in \ac{HCI}~\cite{lazar2017, muller2014} and using the formula by~\citet{krejcie1970}. 
As this study is exploratory, we opted for a confidence interval of $\text{CI}{=}95\%$ and a margin of error $\epsilon{=}10\%$. Assuming ${\approx}3$ billion players worldwide~\cite{gilbert2020}, the recommended sample size was $97$ participants. 
Hence, accounting for $10\%$ data loss, we targeted around $120$ participants to fill in the survey. We advertised the survey on social media (Twitter, Facebook, LinkedIn, Mastodon, Reddit (/r/SampleSize/), and via the personal networks of the authors, and accepted responses from August 18th, 2023 to August 27th, 2023. Average response time was \MeanSd{17.56}{7.44}~\si{\minute}, ranging from $5.56$ to $40.16$~\si{\minute}. 
The research protocol was approved by <removed for review> ethics board, and informed consent was provided at the beginning of the survey.

\subsection{Data Analysis}
The survey received $116$ responses, out of which $111$ were complete and included in the analysis. 
The remaining five responses were not complete and, therefore, discarded. Data were analyzed using Jupyter Notebooks with Python $3.11$ using the Python data stack, including pandas~\cite{reback2020}, numpy~\cite{harris2020}, and pingouin-stats~\cite{vallat2018} packages. Prior to analysis, the data was cleaned, i.e., inverted negative scales, and re-coded in a human-readable format. We provide descriptive statistics for all assessed items, and provide violin plots to facilitate visual inspection of our data as well as distribution thereof. Qualitative data were coded deductively through thematic analysis~\cite{blandford2016} in accordance with the categories in which they were obtained from respondents to further explain quantitative ratings, mapping onto the original themes derived from Phase $1$ (see Section \ref{sec:interviews}).


\subsection{Results}
\begin{table*}[ht]
\caption{Descriptive statistics and Welch's t-tests comparing the sense of achievement, closure, and satisfaction of the positive and negative exits}
\label{tab:pos_neg_exp_feeling}
\begin{tabular}{@{}lrrrrrr@{}}
\toprule
             & Positive Exp. M (SD) & Negative Exp M (SD) & t      & df      & p               & Cohen's d \\ \midrule
Achievement  & $6.069$ $(1.151)$        & $3.062$ $(1.906)$       & $12.49$  & $125.079$ & {\textless}0.001 & 1.961     \\
Closure      & $5.158$ $(1.617)$        & $2.506$ $(1.776)$       & $10.418$ & $163.823$ & \textless 0.001 & 1.570     \\
Satisfaction & $6.198$ $(1.123)$        & $2.728$ $(1.891)$       & $14.582$ & $123.719$ & \textless 0.001 & 2.293     \\ \bottomrule
\end{tabular}
\Description{Table showing the descriptive statistics and Welch's t-tests comparing the sense of achievement, closure, and satisfaction of the positive and negative exits. Means and standard deviations are provided in columns 1 and 2, respectively. Column three shows the t-values. In column four, the Welch-adjusted degrees of freedom are presented. Column five holds the p-values, and the last, most right column shows Cohen's d effect sizes.}
\end{table*}

In this section, we first provide an overview of the respondents. We then structure the results around the three main themes derived in Phase $1$ (see Section~\ref{sec:interviews}):
\begin{enumerate*}[label=(\arabic*)]
    \item[\nameref{subsubsec:interview_theme_1}:] Positive disengagement experiences are associated with satisfaction and closure, while negative disengagement experiences come with feelings of rage and regret,
    \item[\nameref{subsubsec:interview_theme_2}:] disengagement is a complex process that spans players, games, and their environment, and
    \item[\nameref{subsubsec:interview_theme_3}:] players' agency is a central element of positive disengagement
\end{enumerate*}.

\subsubsection{Characteristics of Respondents}
Respondents were \MeanSd{30.50}{7.315} years old with ages ranging from $17{-}59$. 
The sample consisted of $78$ men, $27$ women, $3$ non-binary persons, and $3$ respondents who preferred to self-describe their gender as genderfluid (\n{2}) and Demigirl (\n{1}). 
The participants resided in $20$ different counties with top five countries: Germany (\n{47}), USA (\n{11}), Portugal (\n{9}), Belgium (\n{8}), and Saudi Arabia  (\n{8}). 
Six respondents work in the games industry, and $19$ are game researchers. 
Ninety-nine respondents play games as a hobby, and eight stated that they develop games as a hobby. Considering gaming habits, $41$ play daily, $54$ respondents reported playing a few times a week, and $16$ play once per week or less. 
Average play sessions lasted \MeanSd{122.92}{85.85}~\si{\minute}, ranging from $2$ to $480$~\si{\minute}. In terms of devices, \n{71} play on desktop \acp{PC}, \n{61} on consoles, \n{48} on smartphones, \n{37} on laptops, \n{13} on tablets, \n{11} in \ac{VR}, and \n{3} stated the play other devices: Steam Deck, Physical/tabletop games, and Handheld Consoles respectively. 
Most participants play at home (\n{111}), $32$ also play during commutes, nine play at work, and six play at school or university. Considering the time when participants typically play, $20$ play in the morning, $45$ throughout the day, $88$ in the evening, and $72$ play late at night. 
In terms of mode of play, our sample is inclined towards single-player games (\MeanSd{2.57}{1.36}, Very Frequently - Frequently, 7pt scale, \tdf{110}{-5.139}, \psig[<]{0.001}, \cohensD{0.87}) over multiplayer games (\MeanSd{3.92}{1.71}, Frequently - Occasionally, 7pt scale). 
As an indicator of the participants' play preferences, we asked them to name three of their favorite games (which many participants noted they found difficult to choose). 
After cleaning, the sample consisted of $213$ unique game titles. 
To collect further information about the game, we searched each game title on IGDB\footnote{\url{https://www.igdb.com}} and extracted the total rating, game modes, and genres the games are associated with. 
Overall, the participants played mostly high-quality games with an average rating of \MeanSd{83.88}{9.37} (min: $36$, max: $99$). 
For the game mode, most games allowed for single-player (\n{182}, e.g., Legend of Zelda series~\citeg{nintendoepd2023}), followed by multiplayer (\n{110}, e.g., League of Legends~\citeg{riotgames2009}), co-operate (\n{61}, e.g., Journey~\citeg{chen2012}), \ac{MMO} (\n{13}, Final Fantasy 14~\citeg{squareenix2013}), and battle royal (\n{4}, PUBG~\citeg{pubgstudios2017}). 
The reported games are labeled with $22$ genres with the top five most prominent being Adventure (\n{102}, e.g., The Witcher series~\citeg{cdprojektred2015}),  \acp{RPG} (\n{76}, e.g., Diablo series~\citeg{blizzardentertainment2023}), Indie (\n{44}, e.g., Kerbal space program~\citeg{squad2011}), Strategy (\n{43}, e.g., StarCraft 2~\citeg{blizzardentertainment2010}), and Shooter (\n{42}, e.g. Overwatch 2~\citeg{blizzardentertainment2022}). 

\subsubsection{Respondents' Exit Experiences} 
We examined respondents' personal exit experiences from multiple angles: First, we screened how easy or difficult they experience regulation of their gaming time on a 7-point Likert scale ranging from Strongly disagree ($1$) to Strongly agree ($7$). 
Most participants rated it "somewhat" easy to regulate the gaming time (\MeanSd{4.874}{1.421}). 
Second, we explored individual accounts of exiting play. 
The large majority of respondents have previously had both positive and negative experiences of disengagement from play \n{85}. 
Positive examples largely included those where players had experienced peak moments in gameplay that were associated with achievement and allowed them to exit play with a sense of satisfaction, e.g., \textquote[S026]{\textelp{} it was very rewarding when defeating Ganon and getting the final cutscene. I felt relieved and happy to have saved the "world"}{.} 
However, we also observed more nuanced accounts. 
For example, one respondent pointed out that \textquote[S101]{\textins{t}he game just had a wonderful story, and the world was very alive and full. After exiting, I was happy to know that I could go on next day thinking about all the stuff still to explore}{.} 
Interestingly, data show that positive exit experiences can also be associated with negative emotions, e.g., a sense of loss when players realize that an enjoyable session is about to end, \textquote[S030]{I decided to stop playing because it was getting too late \textelp{}. 
I felt slightly disappointed that it was already time to stop playing... Time flies when you're having fun}{.} 
The majority of negative exit experiences were associated with poor in-game performance and frustration, which were either impacted by high game difficulty or slow gameplay that led to boredom.
For example, instances of failure perceived as unfair triggered negative disengagement experiences, e.g., \textquote[S007]{\textelp{} the lack of any warning or indicators I perceived as unfair. This frustrated me as the city of many hours is now without people, and I had to start over}{.} 
Likewise, one participant highlighted how \textquote[S058]{\textins{t}he game still keeps on riling you on with new quests and tasks, so you think to yourself, "just one more mission". 
It's also kind of leading you on, meaning you get a mission, you finish it, but then there is a plot twist that basically means you did not finish that task and leads to the next}, underscoring the relevance of transparency. 
Quantitative data show that memories of positive exit experiences were strongly associated with a sense of \textit{achievement}, \textit{closure}, and \textit{satisfaction}. 
In contrast, these aspects were less pronounced in negative exit experiences. 
To assess how strong the three feelings were associated with positive and negative disengagement, we conducted paired Welch t-tests with \textit{achievement}, \textit{closure}, and \textit{satisfaction} as dependent variables and positive/negative disengagement as independent variables.
The t-tests confirmed a significant difference with a strong effect (see Table~\ref{tab:pos_neg_exp_feeling}), indicating that achievement, closure, and satisfaction are strongly associated with positive disengagement, while they are less pronounced in negative disengagement.

\begin{figure}[h]
    \centering
    \begin{subfigure}{0.45\textwidth}
        \includegraphics[height=5cm] {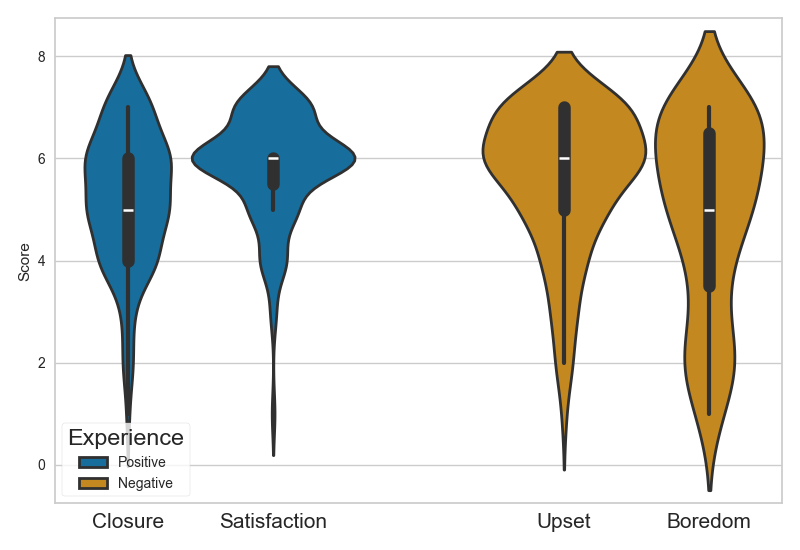}
        \subcaption{Positive and negative Disengagement Experiences: Participants rated how strongly they associate the fulfillment of \emph{Closure} and \emph{Satisfaction} for positive disengagement, and\emph{Upset} and \emph{Boredom} with negative disengagement.}
        \label{fig:disengagement_plot_exp}
        \Description{Violin plots of respondents' ratings of positive and disengagement experiences on a scale of Strongly disagree ($1$) to Strongly agree ($7$). The violins from left to right:  Closure, Satisfaction, Upset, and Boredom.}
    \end{subfigure}
\end{figure}
\begin{figure}[h]\ContinuedFloat
    \begin{subfigure}{0.45\textwidth}    
        \includegraphics[height=5cm]{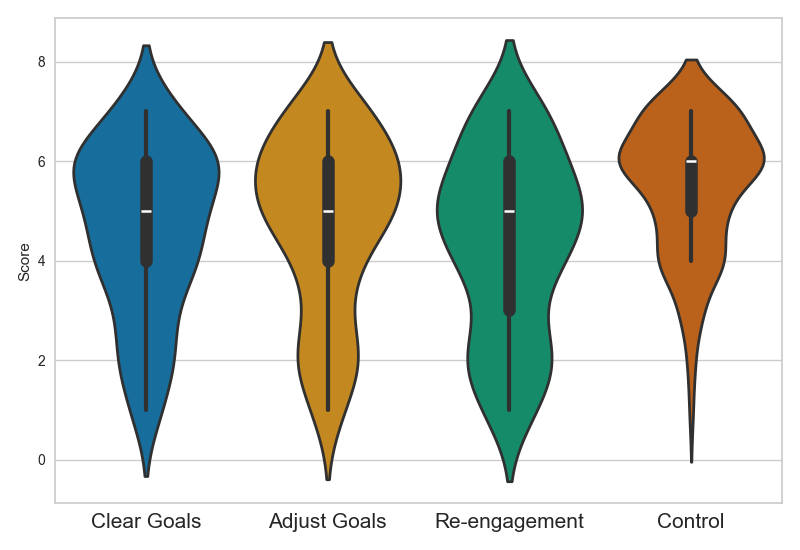}
        \subcaption{Disengagement Process: Participants rated on the relevance of setting \emph{Clear Goals} before a play session, if players \emph{Adjust Goals} for good exits, players' considerations of \emph{Re-engagement} at exits, and importance of being in \emph{Control}.}
        \label{fig:disengagement_plot_process}
        \Description{Violin plots of respondents' ratings of the disengagement process on a scale of 1-7. The violins from left to right: Clear Goals, Adjust Goals, Re-engagement, and Control.}
    \end{subfigure}
\end{figure}
\begin{figure}[h]\ContinuedFloat
    \begin{subfigure}{0.45\textwidth}
        \includegraphics[height=5cm]
        {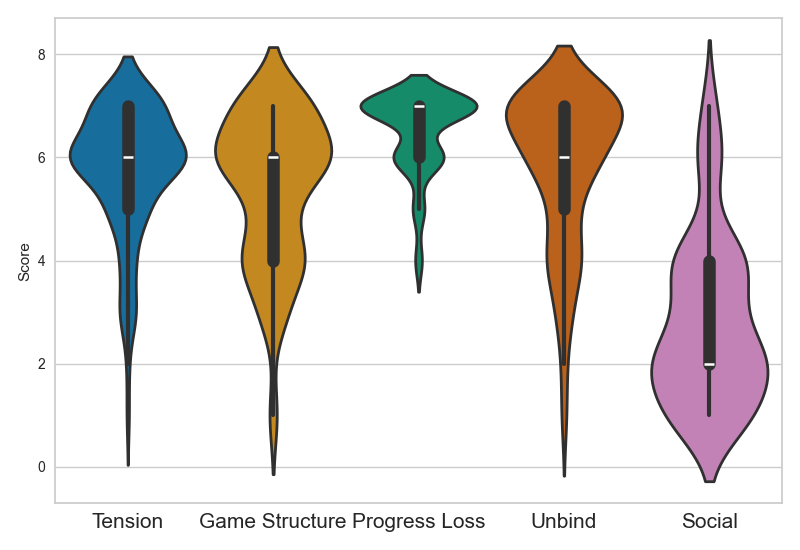}
        \subcaption{Game Characteristics Affecting Disengagement: Participants gave their degree of agreement if low \emph{Tension} in the game, understanding the \emph{Game Structure}, avoidance of \emph{Progress Loss} the game's ability to \emph{Unbind}, and \emph{Social} play eases exits.}
        \label{fig:disengagement_plot_game}
        \Description{Violin plots of respondents' ratings of game characteristics affecting disengagement on a scale of 1-7. The bars from left to right: Tension, Game Structure, Progress Loss, Upset, and Social.}
    \end{subfigure}    
    \caption{Violin plots showing probability densities and box-plots of participants' ratings of characteristics of general disengagement on seven-point Likert scales (1 = strongly disagree, 7 = strongly agree): (\subref{fig:disengagement_plot_exp}) positive and negative disengagement experiences, (\subref{fig:disengagement_plot_process}) the disengagement process, and (\subref{fig:disengagement_plot_game}) game characteristics affecting disengagement.}
    \label{fig:disengagement_plot}
\end{figure}

\subsubsection{Player Perspectives on the Experience of Disengagement}
The respondents agreed that they associate positive disengagement with the sense of closure (\MeanSd{5.198}{1.313}) and satisfaction (\MeanSd{5.829}{1.035}). 
For negative disengagement, there was an agreement with its association with feeling upset (\MeanSd{5.586}{4.883}); however, this was less pronounced for the experience of boredom (\MeanSd{4.883}{1.915}).
The distributions of the ratings are depicted in the violin plots in Figure~\ref{fig:disengagement_plot_exp}, showing a prominent accumulation of the responses around the means and supporting a strong agreement with the statements.
To better contextualize the results, we report the rationale of the responses from the open-ended comments. 

Further elaborating on their experience of closure, several respondents highlighted that they appreciated play sessions that, in wrapping up, also allowed them to look forward to the next. For example, \textquote[S026]{I agree, but not only closure. Also, knowing that you have entered a new area or started a new chapter could give you a positive feeling. You have completed a part, and maybe a new, exciting part awaits you the next time}{.} 
With this anticipation, players may remain mentally engaged with the game, drafting their next step.~\textquote[S029]{Sometimes, but I also like being able to set myself up with a clear next step that I'm excited about for the next time I can play}. 
In contrast, some respondents stated that they were indifferent about closure at the end of a session, and were mostly interested in the ease of re-engagement, e.g., \textquote[S059]{I'm fine as long as I don't feel it will be difficult to pick up where I left off, and that I remain interested to return}. 
With respect to satisfaction, there was strong agreement with its relevance for positive disengagement experiences also in qualitative feedback, e.g.,~\textquote[S096]{\textelp{e}specially after achieving the pre-defined goal of the session}. 
Finally, participants reported positive disengagement when they enacted agency and felt empowered. For instance, S022 commented:, \textquote[S022]{Positive exit experience is when I indeed exited the game at the moment I planned. It makes me a bit proud of my willpower}{.}

In contrast, responses regarding negative disengagement experiences strongly emphasize the role of disappointment and frustration, e.g., \textquote[S002]{Not angry. I mostly feel disappointed, like I was expecting to eat a snack. 
The package is there, but it is empty}{.} 
Disappointment was related to unachieved goals in the game.
Such examples are ~\textquote[S042]{failing at defeating a boss}), expectations that were not satisfied (\textquote[S006]{Yes, especially when I am not able to enjoy the things in the game where the game should be strong in e.g., Baldur's gate 3: roleplay and exploration}), or when the players felt~\textquote[S026]{\textelp{} wasted some time or need to redo some parts of the game}. 
Another frequently mentioned source of negative disengagement experiences is social pressure and expectations of co-players. \textquote[S096]{Sometimes I feel pressured to play longer than I usually would because I don't want to disappoint my playmate}.
\textquote[S60]{Yes, but not always. This does occur, but equally, it might be because I've ran out of time to play. I may not want to stop, but life calls! This isn't angry or upset, but more a reluctance to want to stop - more like disappointment?}{.} 
Finally, further elaborating on the role of boredom in disengagement, several respondents commented that boredom eases the exit from a play session. 
Interestingly, some understood boredom as a~\textquote[S055]{sign \textelp{} to do something else} and that they play~\textquote[S034]{\textelp{} until \textelp{} bored, and stop. It becomes a natural stopping pace}{.} 
This suggests that in some cases, boredom may lead to disengagement that is neutral, which also explains lower agreement scores in our quantitative data.

\subsubsection{The Process of Disengagement}
Concerning the process of disengagement, respondents strongly agreed with the relevance of being in control (\MeanSd{5.477}{1.334}). 
Furthermore, responses demonstrate the relative importance of clear in-game goals (\MeanSd{4.622}{1.701}), support the notion that players dynamically adjust their goals when exiting play sessions (\MeanSd{4.730}{1.783}), and that re-engagement is considered during exits (\MeanSd{4.324}{1.835}). 
These results are also prominently present in the violin plots in Figure~\ref{fig:disengagement_plot_process}. 
While clear goals, adjusting goals, and being in control revolve around the means and indicate strong agreement, re-engagement shows a wide spread of ratings near the neutral rating.

Qualitative feedback reveals extensive support for the notion that being in control - i.e., experiencing agency - is integral for a positive disengagement experience. 
For example, respondents pointed out that \textquote[S085]{\textins{a} clean cut and not losing any progress due to leaving the game, being in control when to end it and being able to do it quickly} was desirable. 
Even when exiting play due to external factors, the element of agency played a role. \textquote[S003]{I agree. I think that is why most of my disengagement experiences are positive. However, even when I'm not the one who decides to stop (e.g., dinner is ready), still I don't think it turns out to be a negative experience, because I allocated this hour or so before dinner and I was expecting to interrupt the play session}. 
Here, we also observed negative experiences, e.g., for some players, such loss of control~\textquote[S013]{\textelp{} can be frustrating to have to stop due to other commitments}, and instances where the lack of agency impacted the choice of game, e.g., \textquote[S002]{\textins{w}ith a small child, I often have to stop whatever I am doing, which means that if she wakes up during the time I usually play, the game session ends and parenting duties resume. I cannot accommodate games that require more commitment (nor I feel right leaving midway)}. 
Regarding the relevance of clear game-related goals to support disengagement, opinions among respondents diverged, with some leveraging their knowledge about the game to support this process. For example, ~\textquote[S061]{Is also a technique I use, but rather with time when to end than with game-related goals}, and~\textquote[S106]{I will sometimes avoid what seems to be the obvious next goal in a game if I think it will entail a greater time commitment than I am prepared for (e.g., delaying a boss fight when I only have half an hour to play)}{.} 
However, others suggested that this strategy was less helpful for them. ~\textquote[S033]{I usually stop my session after a certain amount of time and not some concrete goal. Setting a goal helps me sometimes, but not always. Sometimes, reaching that goal gives me motivation to play some more}{.}
Correspondingly, snoozing the timer was mentioned frequently, e.g.,~\textquote[S072]{If I'm engaged by the game, I often have a hard time quitting it, even if I set goals for myself. Just... one... more... turn... ;)}. 
Overall, qualitative feedback supports the notion of dynamic adjustment of the moment of disengagement, and the relevance of leaving at an adequate moment within gameplay, e.g., ~\textquote[S088]{Totally agree, when I feel this is a good place to take a break, I rather exit my session earlier than having to leave the game in the middle of action}, and \textquote[S060]{This presents a great time to contemplate whether you have time/the capacity to carry on}{.}
Likewise, respondents acknowledged the impact of game features on this process, e.g., 
\textquote[S026]{In the case of games with limited save features, I tend to play enough to reach a save point if I think one might be close. Otherwise as I said, I would quit anyway if it gets too late}{.} With respect to how players plan for re-engagement with the game, we observed many instances in our data, e.g., ~\textquote[S030]{I try to quit at a 'quiet' moment, e.g., right before or after completing a challenge. That way, there is less going on, less to remember when I come back to the game}{.}
Interestingly, one participant stated that to respect the remaining playtime, they change their play style~\textquote[S048]{I occasionally do something slightly different in game if I know I will be quitting soon, so that I will be able to restart from a position that makes sense}{.}
Finally, participants emphasized differences in disengagement processes between single-player and multiplayer games~\textquote[S083]{In single-player games where I can't save anytime, I wait till I can save when I want to exit the game session. In single-player games where I can save anytime, I usually don't have trouble to exit the game session when I want to. In multiplayer games, I don't adjust my in-game goals. I usually exit the session when the round I planned to be the last, is over}{.} Likewise, feedback from respondents also supports the impact of a player's emotional state on their ability to carry out a structured disengagement process, e.g., ~\textquote[S077]{When I am in a good mindset, \textins{I try to exit at a good moment in play}. Usually, that goes out of the window when I have a bad day}{.}

\subsubsection{Game Characteristics Affecting Disengagement}
Considering how the game itself, we asked participants to rate how different aspects of the game itself affect disengagement. 
The respondents expressed strong agreement with the supporting nature of a slower period with low tension in gameplay (\MeanSd{5.748}{1.232}), the relevance of the game structure (\MeanSd{5.360}{1.463}), avoidance of progress loss (\MeanSd{6.495}{0.773}), and the game's ability to unbind the player; i.e., the practical support of exiting play that a game offers (\MeanSd{5.802}{1.500}). In contrast, social play was associated with rather an obstacle to disengagement (\MeanSd{2.840}{1.638}). 
As depicted in Figure~\ref{fig:disengagement_plot_game}, the distributions of the ratings show strong consistencies around the means and indicate a strong agreement with the statements.

Qualitative feedback reveals that players generally appreciate phases of low tension to exit the game and reflect on their decisions, e.g., \textquote[S073]{In single-player games, yes. So when the action is finished, and it's more on the exploration side, it's a good time to save and exit the game}{.} However, slow-paced phases also allow players to change their playstyle and reflect on the gameplay, e.g., \textquote[S060]{I LOVE pacing in games, slower parts allow for reflection/chatting with friends/a chance to grab a drink, for instance. Great for accessibility}{.} Conversely, some players have an ~\textquote[S032]{easier time quitting when the game is about to give me a major challenge}. 
In contrast, disengagement is experienced as more challenging when the structure of the game is unclear. For example, \textquote[S108]{When I discover new games to play, I find it difficult to limit my playing time, as I am unfamiliar with the overall structure and the time required to complete levels, events, etc}{.} Likewise, distractors in the game were detrimental to the ability to exit, e.g., \textquote[S030]{\textelp{g}ames like The Legend of Zelda: Tears of the Kingdom are built around 'distractions' and points of interest to guide players all over the map. This makes it hard to set exit goals}.
However, our data also suggests that very obvious structures can cause players to lose interest in games, e.g.,
\textquote[S050]{But sometimes it makes the game not interesting for me anymore because I "already figured out the process"}{.} Considering features that support player agency throughout disengagement, respondents almost unanimously valued save points as essential for positive disengagement, arguing, \textquote[S034]{\textelp{t}his makes exiting seamless}{.}
Several comments noted the frustration of having~\textquote[S020]{to leave and have to redo your work, so this is a crucial feature, especially if things are going well}. Furthermore, the participants noted the ease of disengagement is dependent~\textquote[S055]{\textelp{} on the amount of progress \textelp{they} will lose}. This was closely related to the importance of being given the tools to exit a session at any time, with respondents highlighting that it would be~\textquote[S026]{user-friendly} if the game allows players to leave~\textquote[S013]{\textelp{} within a reasonable timeframe}, especially for single-player games. A number of respondents commented that multiplayer games are particularly challenging with respect to disengagement and pointed out that, \textquote[S047]{\textelp{s}ome games, like competitive multiplayer games, should not allow you to leave early without punishment}. Again, respondents highlighted the importance of agency with respect to features that regulate playing time, e.g., \textquote[S061]{\textelp{t}he game should not determine how long I play it}. Finally, respondents elaborated on their perspectives on disengagement in the context of multiplayer experiences. Interestingly, many commented that when playing with others, it is often~\textquote[S054]{easier to find a natural `exit' point together}, and frequently teammates \textquote[S060]{\textelp{} allocate a time together to play to or keep an open dialog on when we're heading off, so this makes it easier to set endpoints}. In contrast, one participant commented that \textquote[S098]{\textelp{} it is easier to lose track of time when playing with friends}{}. Likewise, several respondents noted that~\textquote[S062]{social pressure may hold you back in the game~\textelp{}} and \textquote[S059]{\textelp{l}eaving mid-session might be disrespectful to others' time}.

\section{Discussion}
Here, we first provide answers to our research questions, and we then discuss points for reflection for the design of games that address disengagement as a natural part of play. We close with a situation of our findings on player disengagement within wider \ac{HCI} games research.

\subsection{RQ1: How do players experience disengagement from a play session, and what is considered a positive exit experience?}
Our work shows that disengagement experiences are emotional, often involving peak moments in player experience in which gameplay culminates from a structural perspective. For many players, this is associated with a feeling of exhaustion at the time of disengagement, which can be a contributing factor to a player's decision to disengage. This is in line with previous work focusing on continuous engagement~\cite{tyack2021}; in our case, the peak and immediate drop in action and experience are facilitating factors that enable players to disengage. Particularly addressing the question of what players perceive as a positive disengagement experience, our work highlights that those experiences that enable players to exit with a sense of satisfaction and closure are perceived as particularly rewarding. Here, we find that perception of sufficient in-game progress and player performance contribute to satisfaction, and that alignment of a structural end-point in play (e.g., end of a level or mission) with the anticipated duration of play contributes to a sense of closure, all leading to a positive exit experience. Interestingly, our work also shows that such positive exit experiences were in some cases still paired with negative emotions, e.g., a feeling of loss when leaving a game at the end of a rewarding play session, but that players were willing to accept this experience in exchange for time that they perceived as well spent. In contrast, exit experiences deemed negative were those that occurred in response to negative in-game events and those where player agency in terms of managing the process of exiting the play session was hampered, which is an aspect that we discuss in the following section. Finally, we also want to highlight that between the poles of positive and negative disengagement, there exists the experience of neutral disengagement: Our work shows that it is often paired with boredom, yet no frustration, and is an experience of a game slowing down, and the players no longer feeling like they need to engage. This perspective is not yet included in the Engagement Cycle (see Section \ref{subsec:engagement_cycle}), and suggests a need to expand how we categorize (dis)engagement.  

\subsection{RQ2: How do players structure the process of disengaging from a play session, and which characteristics of games help or hinder this process?} 
We find that players indeed approach disengagement as a process, and do so strategically. Our results show that temporary disengagement or exiting a play session is a process comprising multiple steps that extend beyond immediate engagement with the game, including a pre-play planning phase, ongoing in-game adjustment of disengagement goal culminating in a decision to disengage, and in some cases post-play reflection with a plan for re-engagement. Here, the pre-play planning phase includes player considerations regarding their available time and time requirements of specific games (taking into account at which point in play it is possible to achieve closure), highlighting the relevance of games clearly communicating their structure to players. The in-game adjustment of disengagement goals allows players to respond flexibly to gameplay, e.g., by finishing a session slightly earlier or later if conducive to achieving closure. Here, disengagement is impacted by the tools that games provide players with to enact the exit from play. For example, retaining in-game progress is highly relevant to players, and the availability of an opportunity to save the game may shift the point of disengagement. Likewise, our results show that factors external to the immediate interaction between player and game can affect the point of disengagement. For example, the overall state of the player (e.g., whether they are tired) or a social situation (e.g., co-players deciding to exit) may trigger a disengagement decision that diverges from their initial plan. Finally, in some instances, disengagement also included a critical appraisal of the play session or making concrete plans for re-engaging with the game, supporting the notion that engagement is, in fact, cyclical~\cite{obrien2008}.



\subsection{Points for Reflection for Game Design: Understanding and Supporting Player Disengagement}
Our work can support game design through characterization of disengagement experiences, allowing designers to better understand how players experience the final stages of play, and approach the process of disengagement. Here, we summarize the three main points for reflection derived from our work, and illustrate their relevance through positive and negative experiences of play as well as connection with previous research.

\subsubsection{Reflection 1: Valuing the importance of temporary disengagement and understanding its impact on player experience is an opportunity for game design.}
Although many games are designed for disengagement at the final moment of play (i.e., when a game ends), temporary disengagement at the end of a play session should also be regarded as an opportunity for design. While games already support this through the opportunity to save progress, our results show that well-crafted structures of play (e.g., explicit ones such as levels or missions or implicit ones such as boss battles) can further facilitate disengagement in a positive context when clearly communicated to players. Here, save points anchored in the game world can serve as a predefined structure to provide a sense of closure to the players. Additionally, our work provides empirical support for previous considerations for game design by~\citet{bjork2004}, who have included \textit{closure points} in their game design patterns for game sessions and view these as a means of supporting transparency (\textit{predictable consequences}) and a moment in which \textit{transfer of control} to the player can occur. 
In this context, supporting players in building mental models of how gameplay is structured is an opportunity for game design to allow players to pre-empt the required time investment, and to make informed decisions in how they shape their play sessions. With this, transparent game design can contribute to self-determined experiences of disengagement.
For instance, game structure can be made transparent through clear level design with explicit goals or an overview of the upcoming tasks in the game. A positive example is the game world map in the Super Mario World~\citeg{miyamoto1990} series, where players can examine their progression in their current world, and can explore which levels contain a boss fight.



\subsubsection{Reflection 2: Games need to account for player affect and provide exit processes that can be navigated by physically and/or emotionally exhausted players.} 
Our work shows that disengagement is closely related to player affect, demonstrating that the affective dimension articulated by prior work~\cite{schoenau-fog2011} also remains relevant in the context of disengagement.
Here, designers need to be aware of the emotional experience of the last minutes in a gaming session, supporting players at a moment when player experience can be bittersweet, for example, combining the experience of competence with a sense of loss (e.g., winning a boss battle, and then disengaging from play). 
Additionally, our work shows that players often disengage from games when they experience exhaustion, e.g., psychological fatigue after intense periods of play (cf.~\citet{tyack2021} for a discussion of (extra)ordinary player experience) or physical exhaustion in the case of games that require movement-based input. This finding aligns with previous work on spaced practicing (cf. ~\cite{johanson2019, johanson2023}), suggesting that regular breaks from games can improve player performance. This is especially relevant for (semi-)professional competitive play, where players are particularly concerned about spending their time effectively on training. For example, the strategy game \textit{Stronghold}~\citeg{bradbury2001} features a non-player character who pops up regularly and reminds the player to take a break from the game.
Likewise, there were many accounts in our research that suggest players engage with games in their evenings, exiting play when they begin to feel tired. Here, it is important to account for player cognition (which can, for example, be affected by a lack of sleep~\cite{killgore2010}), understanding that the player's ability to perform during the final moments of play may be lower than throughout continuous engagement, but remains nevertheless relevant for a positive exit experience~\cite{gutwin2016}. For example, to account for player fatigue, game design could draw from \ac{DDA}~\cite{hunicke2005} methods to monitor whether players deviate from their regular in-game behavior, and provide opportunities for the player to finish the session.


\subsubsection{Reflection 3: Game design should prioritize player agency by creating games that neither trap players within them, nor push them out.}
While there is a trend to employ (problematic) design strategies to keep players in the game (e.g., see~\cite{zagal2013}), our work shows that players value being able to exit at a point of their choosing. This is in line with general theories of engagement within \ac{HCI} research~\cite{doherty2019} and has straightforward implications for game design that one might perhaps call obvious but that are nevertheless absent from many games: Players need to be handed practical tools to leave the game (e.g., an explicit menu option) and should be able to do so at any time, and with the game being transparent about implications (e.g., whether progress loss - that should be avoided - will occur). Likewise, game design should be aware that player agency is hampered by any strategy that meddles with the point of disengagement, e.g., only offering an option to save progress at certain points of play or artificially limiting playing time through mandatory breaks, risking to \textit{dis}empower players~\cite{vornhagen2023}. 
Here, our work is in line with \citeauthor{howe2017}'s critical contention on practices with free-to-play games~\cite{howe2017} and suggests that current best practices applied by the industry (e.g., time limits, see Section~\ref{subsubsec:disengagement_tools}) directly contradict player preferences, and should not be considered viable approaches to limiting engagement with games under the protection of player agency as they are unlikely to yield positive post-hoc appraisal of play. 


\subsection{Situating Player Disengagement in the Context of HCI Games Research}
While \ac{HCI} games research has long focused on continuous player engagement (see Section~\ref{subsec:disengagement_in_games}), more recent research suggests that there is merit in diversifying the perspective on player experience. Here, our work adds to the body of literature in two ways. First, it highlights that player experience needs to be understood across all stages of play. 
While there is some recognition of the temporal aspect of player experience -- for example,~\citet{tyack2021} discuss \textit{"the temporal dynamics of (extra)ordinary PX"} but remain within the stage of continuous engagement,~\citet{haider2022} address concerns regarding the measurement of \ac{PX} and temporality, and~\citet{cheung2014} examine the first hour of play -- less is known about how players experience the later stages of play. We further contextualize \citeauthor{gutwin2016}'s exploration of Peak-End effects in casual games~\cite{gutwin2016} (see Section \ref{subsubsec:disengagement_unique}): We show that in many instances, players actively seek out experiences that add to their overall player experience toward the end of a gaming session (e.g., events such as final boss battles that facilitate competence and closure), and attempt to structure play sessions accordingly. Second, our findings suggest that another distinct experience of disengagement is associated with boredom that was welcomed by some players toward the end of a gaming session. This finding aligns with games research on \textit{ordinary} player experiences, which demonstrates that less intense moments of play in conjunction with extraordinary experiences can enable players to connect with games in a meaningful way~\cite{tyack2021}. Likewise, slower moments of play are appreciated as a means of recovery from demanding day-to-day tasks~\cite{collins2014}. Here, our findings suggest that many participants engaged with games in the evening, with play naturally ending when they felt tired. Overall, we expect that understanding the experience and process of disengagement from the player's perspective can add nuance to how we approach player engagement. Our work shows that well-supported disengagement contributes to player agency, promoting engagement with games on one's own terms and in a way that facilitates balance~\cite{birk2023}. Additionally, our work shows that players appreciate well-designed disengagement experiences, suggesting that an understanding of disengagement experiences and processes is valuable for designers in an effort to create successful games. However, we also want to highlight the ethical concerns associated with game design that attempts to shape player behavior, where researchers and designers need to ensure transparency in an effort to avoid deceptive design \cite{zagal2013}. 

\section{Limitations and Future Work}
Our work needs to be interpreted in the light of a number of limitations. Most importantly, we studied disengagement experiences in an adult population who are enthusiastic players with no past experiences of problematic play. While we strove to recruit a diverse group of participants at the interview stage, we observed that the participant sample of the online survey predominantly includes men from Western societies and individuals who heavily engage with games; hence, our findings should be interpreted against this backdrop, and we encourage the further exploration of exit experiences within different groups of players (e.g., younger players, women, and non-binary people, or people who only casually play games). Further, this work is exploratory and predominantly qualitative. While this approach is a first step in the investigation of player disengagement, there would also be merit in the development of a measure to quantitatively study disengagement experiences as part of \ac{GUR}. This work also highlights a lack of clear language in the research community to describe fulfilling player experiences, distinguishing achievement, closure, and satisfaction. Here, our work can provide initial directions for future instrument development.

The work presented spanned single- and multiplayer experiences. However, the focus of this research is on disengagement in single-player settings. While we gained some insights into the dynamics of shared play in the context of disengagement, our work shows that disengagement in multiplayer settings is complex and less transparent as different social dynamics influence the players' freedom to disengage. Hence, future work should explicitly examine in more detail how relationships with other players and their specific nature affect the disengagement process, for example, in the context of \acp{MMORPG} or team-based eSports titles. 
Likewise, one may expect disengagement to vary across genres, which was not specifically addressed in this work. Therefore, future work should put effort into characterizing disengagement in different game genres.




\section{Conclusion}
While the games research community has developed substantial knowledge on designing engaging experiences that draw players in, the nature of player disengagement is less clear. Our work addresses this gap by examining the process of disengagement as a natural part of play, taking the first step toward a more holistic perspective on player experience from a procedural perspective. Not just focusing on the first but also on the very last moments in a play session allows us to characterize the experiences that players have when exiting games and puts us in a position to deconstruct how elements and features of games can support or hinder the process of ending play. This new understanding of disengagement as an emotional yet necessary component of play enables our community to develop novel research approaches that address the dedicated design for disengagement, striving to improve the overall player experience by enabling players to adequately exit play sessions. This is not just relevant to support the development of high-quality playful experiences but can also be interpreted as a commentary on ongoing conversations in our and other research communities surrounding deceptive patterns~\cite{zagal2013} and behavioral design~\cite{birk2023}: While the discourse on harmful design strategies is, of course, necessary, we also believe there is a constructive avenue in which our research community approaches game design in conjunction with players, striving to create products that allow all of us to engage in a self-determined way through the provision of games that prioritize player agency at all stages of play.





\begin{acks}
We would like to thank all participants in our interviews and survey for their time, and for sharing their thoughtful insights into how they disengage from play sessions.
\end{acks}



\renewcommand{\bibnumfmt}[1]{[#1]}%
\bibliographystyle{ACM-Reference-Format}
\bibliography{Disengagement}


\renewcommand{\bibnumfmt}[1]{[\citegameprefix#1]}%
\bibliographystylegame{ACM-Reference-Format}
\bibliographygame{ludography}






\end{document}